\documentclass{INTERSPEECH2023}

\interspeechcameraready 
\usepackage{xcolor}
\usepackage{caption}
\usepackage{subcaption}
\usepackage{textcomp}

\title{Real-Time Joint Personalized Speech Enhancement and\\Acoustic Echo Cancellation}
\name{Sefik Emre Eskimez, Takuya Yoshioka, Alex Ju, Min Tang, Tanel P{\"a}rnamaa, Huaming Wang}

\address{Microsoft, Redmond, WA, USA}
\email{\{seeskime, tayoshio, aleju, mintang, taparnam, huawang\}@microsoft.com}

\begin{document}

\maketitle
 
\begin{abstract}
Personalized speech enhancement (PSE) is a real-time SE approach utilizing a speaker embedding of a target person to remove background noise, reverberation, and interfering voices. To deploy a PSE model for full duplex communications, the model must be combined with acoustic echo cancellation (AEC), although such a combination has been less explored. This paper proposes a series of methods that are applicable to various model architectures to develop efficient causal models that can handle the tasks of PSE, AEC, and joint PSE-AEC. We present extensive evaluation results using both simulated data and real recordings, covering various acoustic conditions and evaluation metrics. The results show the effectiveness of the proposed methods for two different model architectures. Our best joint PSE-AEC model comes close to the expert models optimized for individual tasks of PSE and AEC in their respective scenarios and significantly outperforms the expert models for the combined PSE-AEC task.
\end{abstract}

\section{Introduction}
Many institutions have adopted hybrid/remote work using online communication tools in response to the COVID-19 pandemic. This has become a norm as the world transitions to the post-pandemic era. These tools have also become important for connecting family members and friends. However, online communications can easily be disrupted by unwanted interference, including other speakers in the room, background noises, and acoustic echoes. To address this challenge, most online communication tools have adopted real-time speech enhancement (SE) and acoustic echo cancellation (AEC) methods. 

Existing SE, AEC, and joint SE-AEC models are limited as they are usually trained to preserve all human voices present in the near end. Therefore, voices from other speakers in the same environment, i.e., interfering speakers, can ``leak" into the outbound signal, annoying the calls and compromising privacy. 

Personalized speech enhancement (PSE) models have attracted attention due to their ability to remove the interfering speech as well as the background noise~\cite{giri2021personalized,Eskimez2022,thakker2022fast}. The PSE systems are conditioned on a cue from the target speaker, usually a speaker embedding vector such as a d-vector. They remove all other speakers in the input audio except for the target speaker and  suppress the background noise and reverberation~\cite{Eskimez2022,thakker2022fast}. 
However, most existing PSE models do not have AEC capability. 
Although it is desirable for a single model to encompass all of PSE, AEC, and joint PSE-AEC tasks~\cite{zhang2022personalized,yu2022neuralecho}, 
achieving this goal under a strict cost requirement remains a challenge. 

This paper proposes a set of methods for developing efficient causal models that can handle the PSE, AEC, and joint PSE-AEC tasks. Our proposed solutions involve incorporating the attention-based align-block as described in~\cite{indenbom2022deep} to improve the AEC performance for real recordings by softly aligning the microphone and far-end signals. Furthermore, we propose to use the speaker embedding vector only in the latter half of the network and introduce a bypass path during training. This encourages the earlier layers to focus solely on echo and near-end noise removal, while allowing the later layers to effectively remove the interfering speakers and residuals from the early layers. This architecture will help mitigate the undesirable target speaker over suppression (TSOS) that arises from excessive reliance on the speaker embedding vector, especially for echo removal. The models are trained with multi-task learning encompassing AEC, PSE, and joint PSE-AEC. Comprehensive evaluations were carried out by using both simulated and real data. We applied our methods to both a recently proposed time-domain low-cost PSE model called end-to-end enhancement network (E3Net)~\cite{thakker2022fast} and STFT-based VoiceFilter-Lite\footnote{We used ``FFT magnitude'' configuration with power-law compression and reconstructed enhanced audios with noisy phase.}~\cite{Wang2020}. The joint models trained with the proposed methods performed comparatively to single-task PSE and AEC models, while maintaining similar computational costs and limited TSOS.

\section{Related work}

We define PSE as the process of extracting a clean target speech signal in real time from a noisy, potentially overlapped signal based on a speaker embedding vector of a target talker. Our evaluation criteria involve both speech quality and automatic speech recognition (ASR) accuracy. Pioneering works include~\cite{zmolikova17_interspeech}, DENet~\cite{wang18d_interspeech}, SpeakerBeam~\cite{delcroix2020improving}, VoiceFilter~\cite{wang2019voicefilter}, and VoiceFilter-Lite~\cite{Wang2020}. 
We aim to develop computationally efficient causal models for communication applications.

Several efficient causal PSE methods were proposed previously. VoiceFilter is an STFT-based system that combines convolutional and recurrent layers, conditioned by the d-vector extracted from an enrollment audio signal. VoiceFilter-Lite~\cite{Wang2020} is a causal, computationally faster version of VoiceFilter. Personalized PercepNet (PPN)~\cite{giri2021personalized} modified the original PercepNet~\cite{valin2020perceptually} by conditioning it on the speaker embedding. Joint Unified PercepNet (UPN)~\cite{wang2023framework} model was proposed to deal with both personalized and general SE scenarios. \cite{Eskimez2022} proposed a personalized deep complex convolution recurrent network (pDCCRN) that outperformed causal VoiceFilter. It also highlighted the target speaker over-suppression (TSOS) problem resulting from the ambiguity in speaker characteristics and proposed a metric to measure the degree of TSOS with mitigation methods. Personalized E3Net was proposed in \cite{thakker2022fast} to efficiently perform PSE in the time domain. The model outperformed larger models such as pDCCRN with a much smaller computational cost. However, none of these PSE models addressed the AEC task. 

The field of AEC technology has shifted from traditional digital signal processing (DSP) to deep learning (DL) methods~\cite{sridhar2021icassp,cutler2021interspeech,cutler2022icassp}. While some studies adopted a DL and DSP hybrid approach~\cite{casebeer2021auto,ma2020acoustic}, others employed DL-only methods~\cite{zhang2022personalized,yu2022neuralecho,indenbom2022deep,braun2022task,sridhar2021icassp,cutler2021interspeech,cutler2022icassp}. An attention-based soft alignment module was proposed in \cite{indenbom2022deep} to deal with the potentially varying time delays between the microphone and far-end signals that are seen in challenging real-world AEC scenarios. 
In \cite{yu2022neuralecho}, an AEC model with optional speaker embedding-based personalization was proposed, although it was not tested for PSE. More recently, a joint PSE-AEC model, namely personalized gated temporal convolutional neural network (pGTCNN), was proposed utilizing the embeddings of the target and/or far-end speakers~\cite{zhang2022personalized}.
However, the evaluation was limited to simulated experiments without considering the TSOS issue. This model is based on a somewhat simple combination of the microphone signal, the far-end signal, and the embedding input, and it would benefit from our proposed methods.

\section{Improvements to joint PSE and AEC}
\begin{table*}[ht!]
  \caption{Experimental results for PSE and PSE-AEC using VCTK data sets. TS1 includes the target, interfering speaker, and noise. TS2 includes the target speaker and noise. TS3 includes only the target speaker. Test sets with postfix '-echo' also included far-end echo signals. All scenarios include reverberation.}
   \vspace{-.8em}
  \centering
 \resizebox{1.0\textwidth}{!}{
\begin{tabular}{llcccclccccccclcccccc}
\hline
\multicolumn{1}{c}{} &  &  & \multicolumn{3}{c}{TS1} &  & \multicolumn{3}{c}{TS1-echo} &  & \multicolumn{3}{c}{TS2} &  & \multicolumn{3}{c}{TS2-echo} &  & \multicolumn{2}{c}{TS3} \\ \cline{4-6} \cline{8-10} \cline{12-14} \cline{16-18} \cline{20-21} 
\multicolumn{1}{c}{} &  &  & WER$\downarrow$ & DNSMOS$\uparrow$ & TSOS$\downarrow$ &  & \multicolumn{1}{l}{WER$\downarrow$} & \multicolumn{1}{l}{DNSMOS$\uparrow$} & \begin{tabular}[c]{@{}c@{}}AECMOS\\ ECHO$\uparrow$\end{tabular} &  & WER$\downarrow$ & DNSMOS$\uparrow$ & TSOS$\downarrow$  &  & \multicolumn{1}{l}{WER$\downarrow$} & \multicolumn{1}{l}{DNSMOS$\uparrow$} & \begin{tabular}[c]{@{}c@{}}AECMOS\\ ECHO$\uparrow$\end{tabular} &  & WER$\downarrow$ & TSOS$\downarrow$ \\ \hline
\multicolumn{21}{c}{Baseline systems} \\ 
No Enhancement                          &  &  & 43.03          & 2.92          & 0             && 54.78          & 1.18          & 2.42          && 13.35          & 2.98          & 0             && 33.10 & 2.0  & 2.37 && 7.12          & 0             \\ 
AlignCruse~\cite{indenbom2022deep} - AEC                              &  &  & 57.10          & 3.27          & 0             && 62.46          & 2.39          & 3.53          && 21.82          & 3.41          & 0             && 37.29 & 2.58 & 3.85 && 7.47          & 0             \\ 
pGTCNN~\cite{zhang2022personalized} - PSE-AEC                              &  &  & 40.12          & 3.42          & 1.48             && 52.29          & 2.61          & 4.17          && 20.14          & 3.59          & 0.63             && 32.60 & 2.86 & 4.32 && 7.95          & 0.56             \\ \hline
\multicolumn{21}{c}{Effects of proposed improvements on VoiceFilter-Lite and E3Net} \\ 
VoiceFilter-Lite~\cite{Wang2020} \\
\hspace{0.4cm} - AEC                                &  &  & 49.32          & 3.24          & 0.47          && 59.57          & 2.51          & 4.05          && 20.52          & 3.44          & 0.14          && 33.99 & 2.82 & 4.23 && 7.90          & 0.02          \\ 
\hspace{0.4cm} - PSE                               &  &  & 40.13          & 3.25          & 1.19          && 53.36          & 2.54          & 3.97          && 19.71          & 3.44          & 0.22          && 34.34 & 2.69 & 4.13 && 7.41          & 0.12          \\
\hspace{0.4cm} - PSE-AEC Na\"ive                &  &  & 48.85          & 3.17          & 3.55          && 55.51          & 2.46          & 3.99          && 22.80          & 3.35          & 1.62          && 34.91 & 2.76 & 4.12 && 8.21          & 1.44          \\
\hspace{0.4cm} - PSE-AEC w/o~SC    &  &  & 41.86          & 3.24          & 1.56          && 52.54          & 2.56          & 4.02          && 20.93          & 3.43          & 0.62          && 34.25 & 2.82 & 4.15 && 7.90          & 0.33          \\
\hspace{0.4cm} - PSE-AEC w/~~~SC     &  &  & 41.63          & 3.25          & 1.66          && 52.21          & 2.59          & 4.04          && 20.39          & 3.44          & 0.48          && 34.18 & 2.84 & 4.14 && 7.69          & 0.37          \\ 
E3Net~\cite{thakker2022fast} \\
\hspace{0.4cm} - AEC                                &  &  & 43.50          & 3.51          & 0.33          && 58.19          & 2.80          & 4.26          && 18.05          & 3.71          & 0.17          && 30.12 & 3.06 & 4.48 && 7.19          & 0.05          \\ 
\hspace{0.4cm} - PSE                               &  &  & 36.91          & 3.49          & 2.15          && 51.40          & 2.69          & 4.23          && 18.35          & 3.74          & 0.33          && 33.51 & 2.95 & 4.41 && 7.54          & 0.32          \\
\hspace{0.4cm} - PSE-AEC Na\"ive                &  &  & 38.70          & 3.47          & 1.54          && 54.06          & 2.64          & 4.24          && 19.76          & 3.70          & 0.96          && 33.85 & 2.87 & 4.39 && 7.82          & 1.38          \\
\hspace{0.4cm} - PSE-AEC w/o~SC    &  &  & 38.96          & 3.43          & 1.45 && 51.88          & 2.63          & 4.27          && 19.43 & 3.68          & 0.51 && 32.70 & 2.91 & 4.46 && 7.46          & 0.06 \\
\hspace{0.4cm} - PSE-AEC w/~~~SC     &  &  & 38.05 & 3.48 & 1.62          && 49.97 & 2.75 & 4.35 && 19.76          & 3.71 & 0.52          && 31.08 & 2.98 & 4.48 && 7.45 & 0.14 \\
\hline
\end{tabular}
}
\vspace{-0.4cm}
\label{tab:vctk}
\end{table*}

\begin{figure}[t!]
    \centering
    \includegraphics[width=\columnwidth]{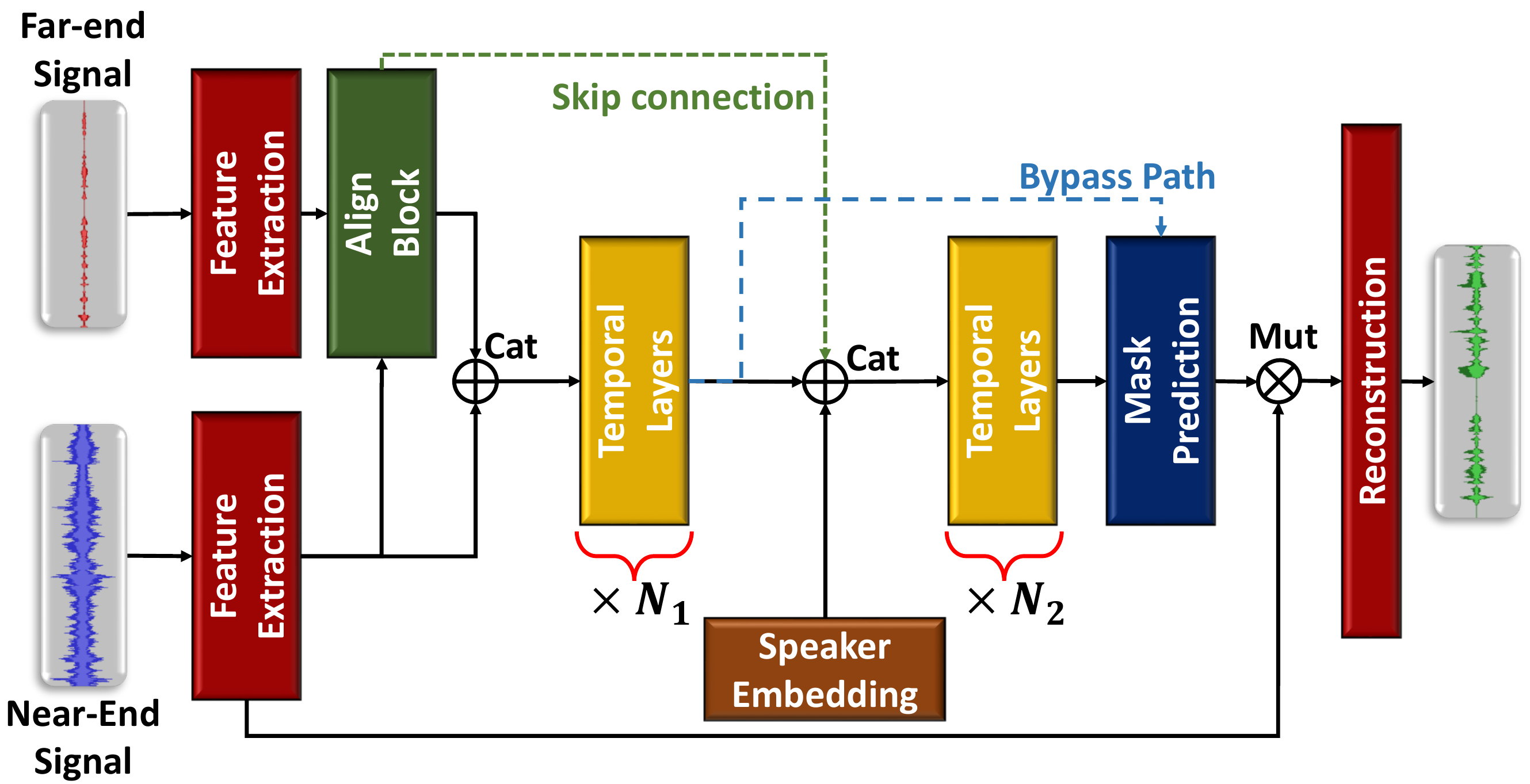}
  \vspace{-0.5cm}
  \caption{E3Net and VoiceFilter-Lite adapted for PSE-AEC task using proposed improvements. `Cat' and `Mut' stand for concatenation and element-wise multiplication, respectively. For E3Net, feature extraction and reconstruction stand for a learnable encoder and decoder, respectively; and temporal layers stand for a projection layer followed by LSTM blocks. For VoiceFilter-Lite, feature extraction and reconstruction stand for STFT and ISTFT, respectively; and temporal layers stand for LSTMs.}
   \vspace{-0.4cm}
  \label{fig:aec_pns_ovr}
\end{figure}

A joint PSE-AEC model must use the microphone signal, far-end signal, and speaker embedding from the enrollment signal as input. One approach is to use a neural network that concatenates the features extracted from the microphone and far-end signals, as seen in \cite{zhang2022personalized}. The speaker embedding can also be included as input to one or more layers of the model. However, this approach is insufficient at handling time delay fluctuations between the microphone and far-end signals. We also found that it suffered from TSOS. Additionally, in practice, some training samples lack speaker labels and enrollment audio files.

We propose a set of methods to address these challenges and apply them to representative time-domain and STFT-domain PSE models, namely E3Net~\cite{thakker2022fast} and STFT-based VoiceFilter-Lite~\cite{Wang2020}. We chose these models as they are computationally efficient, although the proposed methods are applicable to other model architectures. Fig.~\ref{fig:aec_pns_ovr} shows a diagram of PSE-AEC models using the proposed methods. 
Although the diagram and the associated caption provide sufficient information to understand the overall architecture of E3Net and VoiceFilter-Lite, we also refer the reader to \cite{thakker2022fast} and \cite{Wang2020} for the details of the respective models. 
Below, we elaborate only on our key improvements.  

\textbf{Learnable encoder}:
For time-domain models including E3Net, we add a learnable encoder for the far-end signal input. For the microphone signal input, it is beneficial to increase the number of learnable encoder's filters, $F_{mic}$, as shown in \cite{thakker2022fast}. However, for the far-end signal input, a smaller number of learned features, $F_{far}$, can be used without impacting the speech quality since the obtained features are used only for aligning the far-end and microphone signals and estimating the echo intensity for removal.  

\textbf{Align-block}: 
We use the align-block with source-target attention proposed in \cite{indenbom2022deep}, which was shown to be effective in dealing with an unaligned microphone and far-end signals. As input to the align-block, we use the learnable and STFT features for the time-domain and STFT-domain models, respectively. 

\textbf{Bypass path}:
In the original E3Net and VoiceFilter-Lite for PSE, the speaker embedding vector is concatenated with the observed features prior to the first temporal layers. However, our preliminary experiment found that applying this structure to the PSE-AEC task increased TSOS when the far-end signal was not present. In PSE-AEC, both the far-end signal input and the speaker embedding vector may be used as clues for removing the echo signal. We conjecture that the observed TSOS increase was the result of the model becoming over-reliant on the speaker embedding even when the far-end signal provides sufficient clues for the echo removal, which would hurt the model's robustness. To counteract this, we append the speaker embedding vector to the $N_1$th temporal layer's output. For E3Net, another projection layer reduces the dimensionality of the concatenated features to the embedding's dimension and feeds them to the remaining $N_2$ temporal layers. This architectural change is aimed at dedicating the earlier temporal layers to the AEC plus noise removal task. To further encourage the first $N_1$ temporal layers to focus on AEC and noise removal, we add a path bypassing the later $N_2$ temporal layers (the blue dashed line in Fig.~\ref{fig:aec_pns_ovr}) and perform multi-task learning by using AEC mini-batch (see Multi-Task Training below). 

\textbf{Skip connection (SC)}:
Lastly, we append the attention weights from the align-block to the $N_1$th temporal layer's output to help the latter temporal layers adjust the noise suppression behavior based on the presence of the echo signal. 

\textbf{Multi-Task Training}:
Our joint model aims to be as effective as task-specific models for AEC and PSE while outperforming the task-specific models in the PSE-AEC task where the target speaker, interfering speakers, acoustic echo, and near-end noise simultaneously exist in the microphone signal. To this end, we alternate between these three mini-batches with different data configurations in each iteration during training: 
\textbf{AEC mini-batch}---which contains the target speaker, near-end noise, and echo signals. These samples do not include speaker embedding vectors. Hence, the model uses the bypass path to encourage the earlier temporal layers to focus on AEC and noise suppression. \textbf{PSE mini-batch}---which includes the target and interfering speakers as well as noise. The speaker embedding vectors are included, while the far-end signal is all-zero. The full path of the model is trained. The PSE mini-batch helps the model learn the PSE capability. \textbf{PSE-AEC mini-batch}---which includes all signals as well as the speaker embedding vectors. The model uses the full path. This mini-batch helps the model learn to jointly perform PSE and AEC. It also improves the full-path AEC quality by exposing the later temporal layers to training samples with non-zero echo signals.

\section{Experimental results}

To simulate PSE training and validation data, we adopted the same configuration as \cite{Eskimez2022,thakker2022fast}. 
The clean speech utterances were taken from the DNS Challenge data set~\cite{dubey2022icassp}, which is based on LibriVox audio~\cite{kearns2014librivox} and contains 544 hours of speech samples. We used noise samples from AudioSet~\cite{audioset} and Freesound~\cite{freesound}. For each training and validation sample, we randomly placed the target speaker between 0 to 1.3 meters away from the microphone, and the interfering speaker more than 2 meters away by using simulated room impulse responses (RIRs). To improve the model's SE task performance, only half of the samples included interfering speakers. We varied the signal-to-noise ratio (SNR), signal-to-echo ratio (SER), and signal-to-interference ratio (SIR) from 0 to 15 dB, -20 to 40 dB, and from 0 to 10 dB, respectively. The clean speech source for a far-end echo signal was obtained from the DNS Challenge~\cite{dubey2022icassp} and  included singing and emotional speech\footnote{Our training data sources are publicly available and can be reproduced. RIRs can be generated with \url{https://pyroomacoustics.readthedocs.io/}~\cite{scheibler2018pyroomacoustics}.}.

We evaluated our model's performance using simulated long-duration test sets as described in~\cite{Eskimez2022,thakker2022fast}, based on the voice cloning toolkit (VCTK) corpus~\cite{veauxd2017superseded}. These test sets encompass five important scenarios: \textbf{TS1}: target speaker + interfering speaker + noise, \textbf{TS1-echo}: TS1 + echo, \textbf{TS2}: target speaker + noise, \textbf{TS2-echo}: TS2 + echo, and \textbf{TS3}: target speaker only. TS1 and TS2 evaluate the model performance in the PSE and SE scenarios, respectively, while TS3 is used to assess the target speech quality degradation. TS1-echo is the most challenging test set, including all possible signals, and it measures the PSE-AEC performance. TS2-echo evaluates the AEC capability under noise. The SNR and SER were varied from 0 to 15 dB, and the SIR was varied from 0 to 10 dB. VCTK was used as the clean speech source. To create a single long-duration file per speaker, the files from the same speaker were stitched together. Therefore, the acoustic conditions (interfering speaker, noise, RIR, SNR, and SIR) changed every few seconds, making the evaluation challenging. The average duration of the individual test samples was 27.5 minutes. Each test set contains 109 speakers, totaling around 50 hours of data per test set.

\begin{table*}[ht!]
  \caption{Computational complexities of different models and their results for real recording test sets. Real time factor (RTF) was measured on Intel$^{\text{\textregistered}}$ Xeon$^{\text{\textregistered}}$ W-2133 CPU@3.60GHz with a single-thread configuration by taking averages over 1000 independent runs.}
  \vspace{-.8em}
  \centering
 \resizebox{0.88\textwidth}{!}{
\begin{tabular}{llcccccccccccc} \hline
&  & \multicolumn{2}{c}{\textbf{Complexity}}                  &  & \multicolumn{3}{c}{\textbf{DNS Challenge v4 Blind Test Set}} &  & \multicolumn{2}{c}{\textbf{AEC Challenge FST}}                &  & \multicolumn{2}{c}{\textbf{AEC Challenge DT}}                                                                  \\ \cline{3-4} \cline{6-8} \cline{10-11} \cline{13-14} 
&  & \begin{tabular}[c]{@{}c@{}}Parameters \\ (millions)\end{tabular} & RTF &  & SIG$\uparrow$          & BAK$\uparrow$          & OVR$\uparrow$          &  & \begin{tabular}[c]{@{}c@{}}AECMOS \\ ECHO$\uparrow$\end{tabular} & ERLE$\uparrow$ &  & \begin{tabular}[c]{@{}c@{}}AECMOS \\ ECHO$\uparrow$\end{tabular} & \begin{tabular}[c]{@{}c@{}}AECMOS \\ DEG$\uparrow$\end{tabular} \\ \hline
\multicolumn{14}{c}{Baseline systems} \\ 
No Enhancement & & \multicolumn{1}{c}{-} & \multicolumn{1}{c}{-} &  & 3.71 & 2.17 & 2.40 & & 1.98 & 0.0  &  & 1.81 & 4.11 \\ 
AlignCruse~\cite{indenbom2022deep} - AEC & & 0.45 & 0.056 &  & 3.58 & 3.85 & 3.18 &  & 4.12  & 44.40 & & 4.34 & 3.91 \\ 
pGTCNN~\cite{zhang2022personalized} - PSE-AEC & & 5.33 & 0.229 &  & 3.42 & 4.04 & 3.13 &  & 4.39  & 48.27 & & 4.58 & 3.69 \\ \hline
\multicolumn{14}{c}{Effects of proposed improvements on VoiceFilter-Lite and E3Net} \\ 
VoiceFilter-Lite~\cite{Wang2020} \\
\hspace{0.4cm} - AEC & & 8.56 & 0.143 & & 3.43 & 3.86 & 3.07 & & 4.46 & 48.97 & & 4.33 & 3.62 \\ 
\hspace{0.4cm} - PSE & & 8.03 & 0.134 & & 3.44 & 3.91 & 3.11 & & 2.71 & 18.03 & & 4.11 & 3.73 \\ 
\hspace{0.4cm} - PSE-AEC Na\"ive & & 8.36 & 0.138 & & 3.25 & 3.87 & 2.92 & & 4.43 & 45.41 & & 4.26 & 3.24 \\ 
\hspace{0.4cm} - PSE-AEC w/o SC & & 8.36 & 0.142 & & 3.45 & 3.89 & 3.09 & & 4.44 & 45.63 & & 4.31 & 3.69 \\ 
\hspace{0.4cm} - PSE-AEC w/ SC & & 8.56 & 0.143 & & 3.49 & 3.88 & 3.10 & & 4.42 & 45.03 & & 4.32 & 3.71 \\  
E3Net~\cite{thakker2022fast} \\
\hspace{0.4cm} - AEC & & 3.28 & 0.054 & & 3.52 & 4.09 & 3.24 & & 4.47 & 47.91 & & 4.65 & 3.97 \\ 
\hspace{0.4cm} - PSE & & 3.17 & 0.050 & & 3.53 & 4.07 & 3.24 & & 2.20 & 11.42 & & 4.33 & 4.04 \\ 
\hspace{0.4cm} - PSE-AEC Na\"ive & & 3.23 & 0.054 & & 3.49 & 4.06 & 3.19 & & 4.39 & 45.38 & & 4.61 & 3.46 \\
\hspace{0.4cm} - PSE-AEC w/o SC & & 3.27 & 0.054 & & 3.45 & 4.08 & 3.16 & & 4.45 & 49.06 & & 4.61 & 3.92 \\
\hspace{0.4cm} - PSE-AEC w/ SC & & 3.28 & 0.054 & & 3.52 & 4.09 & 3.23 & & 4.41 & 46.78 & & 4.62 & 3.98 \\ \hline
\end{tabular}
}
 \label{tab:real}
\vspace{-0.3cm}
\end{table*}

In addition, we tested the models with real recordings. For the PSE performance evaluation, we utilized the DNS Challenge personalized track~\cite{dubey2022icassp} blind test set, comprising 859 real recordings, 121 of which include interfering speakers. As for AEC, we used the blind test set of the ICASSP 2022 AEC Challenge~\cite{cutler2022icassp}. This test set consists of 600 real-world recordings of 30-45 seconds each, which are split equally into far-end single-talk (FST) and double-talk (DT) cases. Furthermore, this test set includes challenging scenarios commonly found in real online meetings: long and varying delays between microphone and far-end signals, loudspeaker and microphone distortions, non-stationary noise, echo path changes due to moving near-end speakers, audio DSP processing artifacts, and gain variations~\cite{cutler2022icassp}.

\subsection{Evaluation metrics and implementation details}

We evaluated the PSE performance using ASR quality, speech quality, and TSOS. To measure the speech quality, we used DNSMOS P.835~\cite{reddy2021dnsmos}, which is a non-intrusive neural network-based mean opinion score (MOS) estimator and can accurately predict subjective quality ratings. Microsoft's internal ASR model was used to obtain the word error rate (WER). Moreover, we calculated the target speaker over suppression metric from~\cite{Eskimez2022}. We measured the AEC quality using AECMOS~\cite{purin2022aecmos}, which is a neural network MOS estimator similar to DNSMOS, although it was specifically designed for measuring the echo removal quality (AECMOS ECHO) and signal degradation (AECMOS DEG). For the FST scenario, echo return loss enhancement (ERLE) was also used. For the real data without clean reference signals, namely the DNS and AEC Challenge test sets, DNSMOS and AECMOS were used, respectively.

Both training and validation samples were 20 seconds long. For E3Net, we set the $f_{emb}$ and $f_{emb-hid}$ to 128 and 768, respectively. The input and hidden dimensions of the LSTM blocks were the same and set to $f_{emb}$. The learnable encoder and decoder window (filter size) and hop size (stride) were 20 ms (320) and 10 ms (160), respectively. The dimensions of the learnable encoders for the microphone and far-end signals, $F_{mic}$ and $F_{far}$, were 2048 and 256, respectively. The number of the LSTM blocks, $N_1$ and $N_2$ were both 2, totaling 4 LSTM blocks as with the original E3Net. The baseline E3Net training followed the aforementioned parameters and used 4 LSTM blocks. For all VoiceFilter-Lite models, we employed 4 LSTM layers with 512 hidden dimensions and split them equally, as with E3Net. We used power-law compressed STFT magnitudes as input, where STFT parameters followed the same window size and hop size as E3Net. We trained all the models with the power-law compressed phase-aware (PLCPA) loss function (see Eq. (1) of \cite{Eskimez2022}). We employed the pre-trained Res2Net speaker embedding model of~\cite{zhou2021resnext}.

\subsection{Baseline models for individual tasks}

We used AlignCruse~\cite{indenbom2022deep} as our AEC baseline model because of its promising AEC results and low computational requirements. The model is based on STFT input and consists of a 2D convolutional encoder and decoder pair, a recurrent bottleneck block, and an align-block. We used the code and pre-trained model provided by the authors, which outperformed the one reported in the original paper~\cite{indenbom2022deep}. Additionally, we trained E3Net and VoiceFilter-Lite models specifically for the AEC task by removing the d-vector input. Except for the removed d-vector input, these models had the same architectures as the E3Net and VoiceFilter-Lite models for PSE-AEC, respectively. 
It should be noted that some training noise files contained human speech, which may have provided the AEC models with a modest interfering speaker suppression capability.

Meanwhile, as our PSE baseline, we employed a personalized E3Net model and a VoiceFilter-Lite model using 4 LSTM layers. These models were not equipped with the far-end signal input. Note that this E3Net-based PSE model used different configurations than the one used in the previous paper~\cite{thakker2022fast} to ensure a fair comparison with our PSE-AEC E3Net model. 

We also trained PSE-AEC E3Net and VoiceFilter-Lite Na\"ive models to examine the impact of the proposed training and modeling methods. The E3Net Na\"ive model used a learnable encoder for both the microphone and far-end signals and concatenated these features and the speaker embedding vector before the projection layer. For VoiceFilter-Lite, the STFT features of both the microphone and far-end signals were concatenated and coupled with the speaker embedding before they were fed to the first LSTM layer. The Na\"ive models did not contain the align-block, the proposed bypass path, and, in the case of E3Net, the second projection layer. Finally, we trained pGTCNN using the same parameters described in~\cite{zhang2022personalized} to compare our models with a PSE-AEC model that is computationally more expensive. 

\subsection{Results and discussions}

Tables~\ref{tab:vctk} and~\ref{tab:real} show all experimental results for the simulated and real-recording test sets, respectively. 
In the lower half of both tables, we show the results of AEC-only, PSE-only, and three PSE-AEC configurations for both VoiceFilter-Lite and E3Net models. 
The simulation experimental results in Table \ref{tab:vctk} show that the PSE-AEC-Na\"ive configurations consistently produced higher TSOS for both model types. The real-recording test results in Table~\ref{tab:real} also show that the Na\"ive configuration yielded lower AECMOS DEG scores than the other models, suggesting high distortion for the double-talk case. 
Using the skip connection slightly improved the DNSMOS and AECMOS DEG at the cost of FST performance for the real-world recordings and produced lower WERs for the simulated ones, especially for TS1-echo and TS2-echo. 
Finally,  
comparing the PSE-AEC with the proposed improvements and the AEC-only and PSE-only configurations, 
we can see that the joint model achieved good performance for all conditions with both VoiceFilter-Lite and E3Net
at the expense of slight degradations in the test sets that these expert models were optimized for (TS1 for PSE-only and TS2-echo for AEC-only). 
These results demonstrate the effectiveness of the proposed PSE-AEC model improvements for the two model types examined. 

Tables~\ref{tab:vctk} and~\ref{tab:real} also show three baseline results in the upper half.  
The joint PSE-AEC model with E3Net outperformed the recently proposed pGTCNN model for all conditions despite the less computational requirement. 
Also, it outperformed the AEC model based on AlignCruse in terms of AEC performance with similar computational cost. 
These results show the competitiveness of the PSE-AEC model enhanced with the proposed improvements compared with the existing real-time models. 

Finally, to further validate our experimental results, 
we compared the best PSE-AEC E3Net model with recently proposed lightweight PSE models, Personalized PercepNet~\cite{giri2021personalized} and 
UPN-AE-0.9~\cite{wang2023framework}. Since these models were benchmarked with the DNS Challenge v4 development set by using pDNSMOS p.835, 
we also measured our model's performance under the same condition. 
Our model achieved SIG, BAK, and OVR scores of 3.491, 3.982, and 3.096, respectively, and it outperformed both Personalized PercepNet (3.427/3.659/2.880 in SIG/BAK/OVR) and UPN-AE-0.9 (3.454/3.607/2.877 in SIG/BAK/OVR).

\section{Conclusions}

We described a set of methods to improve causal joint PSE-AEC models. Our proposed improvements include adding a new learnable encoder for the far-end signal for time-domain models, an align-block for enhanced AEC robustness, a bypass path that enables multi-task learning, and a skip connection. Our comprehensive evaluation using E3Net and VoiceFilter-Lite showed that the models obtained using the proposed methods were effective in all tasks of PSE, AEC, and PSE-AEC. 

\newpage

\bibliographystyle{IEEEtran}
\bibliography{refs}

\end{document}